\documentclass{PoS}
\pdfoutput=1

\usepackage{amsmath,amssymb,url}
\usepackage{graphicx,subfigure}
\usepackage[euler-hat-accent,euler-digits]{eulervm}
\usepackage{braket,euscript,enumitem}
\usepackage{fancyvrb}

\usepackage{amsfonts}

\usepackage{xcolor}
\usepackage[T1,T5]{fontenc}

\newcommand{\p}{\partial}

\title{BRST invariant $d=2$ condensates in Gribov-Zwanziger theory}

\ShortTitle{BRST invariant d=2 condensates in  Gribov-Zwanziger  theory}

\author{\speaker{Caroline Felix}\thanks{Fellowship from CNPq, Brazil.}\\
        KU Leuven Campus Kortrijk -- Kulak, Department of Physics, Etienne Sabbelaan 53 bus 7657, 8500 Kortrijk, Belgium.\\
        E-mail: \email{caroline.felix@kuleuven.be}}

\author{David Dudal\\
        KU Leuven Campus Kortrijk -- Kulak, Department of Physics, Etienne Sabbelaan 53 bus 7657, 8500 Kortrijk, Belgium.\\
        E-mail: \email{david.dudal@kuleuven.be }}

\author{Leticia Palhares\\
		Instituto de F\'isica Te\'orica, Rua S\~ao Francisco Xavier 524, 20550-013, Maracan\~a, Rio de Janeiro, Brasil\\
		E-mail: \email{leticia.palhares@uerj.br}}

\author{Fran{\c c}ois Rondeau\\
		Ecole Normale Sup\'erieure Paris-Saclay, Avenue du Pr\'esident Wilson 61, 94235 Cachan Cedex, France. \\
	E-mail: \email{francois.rondeau@ens-cachan.fr }}

\author{David Vercauteren\\
		 Duy T\^an University, Institute of Research and Development, P809, 3 Quang Trung, {\fontencoding{T5}\selectfont H\h ai Ch\^au, \DJ \`a N\~\abreve ng}, Vietnam.\\
		E-mail: \email{vercauterendavid@dtu.edu.vn }}

\abstract{In this proceeding, $SU(N)$ Yang-Mills theory is quantized in the linear covariant gauges, while taking into account the issue of Gribov copies and we construct the one-loop effective potential for a set of mass dimension 2 condensates, including the Gribov parameter, that refines the infrared region of the Gribov-Zwanziger theory, whilst respecting renormalization group invariance and BRST symmetry.}

\FullConference{XIII Quark Confinement and the Hadron Spectrum - Confinement2018\\
		31 July - 6 August 2018\\
		Maynooth University, Ireland}

\begin{document}

\section{Introduction}

The analytical explanation of quark and gluon confinement has been a big challenge in recent decades. At low temperatures (or in the infrared regime, IR), where confinement happens, the coupling constant $g^2$ is large, hence the perturbative formalism can not be used in this regime.  This research main focus has been on the gluon propagator and also ghost propagator in the IR. The gluon propagator is suppressed to a nonvanishing value at zero momentum violating reflection positivity and the ghost propagator is not enhanced
at large volume according to the lattice data \cite{Dudal:2018cli}. A possible analytical explanation for this behavior is obtained
through adding dimension 2 condensates to the Gribov-Zwanziger (GZ) formalism, yielding to so-called Refined Gribov-Zwanziger (RGZ) framework \cite{Dudal:2007cw,Dudal:2008sp} that fits the lattice data quite well \cite{Dudal:2018cli}.

In this present study, the analysis of a non-trivial minimum of the effective action, which leads us to a dynamical transformation of the GZ action into the RGZ action, has been done in the presence of the $\Braket{A^h A^h}$ and $ \Braket{\bar{\varphi}\varphi}$ condensates at one-loop following earlier steps of \cite{Dudal:2011gd}, suitably generalized to respect BRST invariance following recent developments by some of us in the field.

In this proceeding, we highlight a few steps, a comprehensive paper will be presented elsewhere.

\section{The Gribov-Zwanziger action in the linear covariant gauge}\label{GZ action}
In the IR region, the Gribov copies appear. Since the coupling constant $g^2$ is large, these copies
can not be eliminated \cite{Gribov:1977wm}.  A way to work around this problem is to restrict the functional integral to a specific region $\Omega$ in field space, a solution proposed by Gribov using the Landau gauge \cite{Gribov:1977wm}. Moreover, this solution given by Gribov can be generalized to linear covariant gauge  \cite{Capri:2015ixa}:
 \begin{equation}
\Omega=\{A_{\mu}^a;\  \partial_\mu A_{\mu}^a=i\alpha b^a,\qquad \mathcal{M}^{ab}(A^h)=-\partial_\mu D_\mu^{ab}(A^h)>0 \}.
\label{GR}
\end{equation}
whereby the Hermitian Faddeev-Popov-related operator, $\mathcal{M}^{ab}(A^h)=-\delta^{ab}\partial^2+gf^{abc}(A^h)_{\mu}^{c}\partial_{\mu}$, is positive. In \eqref{GR}, $A_\mu^h$ is a non-local power series in the gauge field, gotten from the minimization of the functional $f_A[u]$ along the gauge
orbit of $A_{\mu }$ \cite{Dell'Antonio:1989jn,vanBaal:1991zw,Lavelle:1995ty},
\begin{eqnarray}
f_A[u] &\equiv &\min_{\{u\}}\mathrm{Tr}\int d^{d}x\,A_{\mu
}^{u}A_{\mu }^{u},
\nonumber \\
A_{\mu }^{u} &=&u^{\dagger }A_{\mu }u+\frac{i}{g}u^{\dagger }\partial _{\mu
}u.  \label{Aminn0}
\end{eqnarray}
A local minimum is found and given by
\begin{eqnarray}
A_{\mu }^{h} &=&\left( \delta _{\mu \nu }-\frac{\partial _{\mu }\partial
_{\nu }}{\partial ^{2}}\right) \phi _{\nu }\;,  \qquad  \partial_\mu A^h_\mu= 0 \;, \nonumber \\
\phi _{\nu } &=&A_{\nu }-ig\left[ \frac{1}{\partial ^{2}}\partial A,A_{\nu
}\right] +\frac{ig}{2}\left[ \frac{1}{\partial ^{2}}\partial A,\partial
_{\nu }\frac{1}{\partial ^{2}}\partial A\right] +O(A^{3}).  \label{min0}
\end{eqnarray}
Here we highlight that $A_\mu^h$ is gauge invariant  order by order \cite{Capri:2015ixa}.
The field $A_\mu^h$ can be localized by introducing an auxiliary Stueckelberg field $\xi^a$ \cite{Capri:2015ixa,Fiorentini:2016rwx},
\begin{equation}
 A^{h}_{\mu}=(A^{h})^{a}_{\mu}T^{a}=h^{\dagger}A^{a}_{\mu}T^{a}h+\frac{i}{g}\,h^{\dagger}\partial_{\mu}h,    \label{st}
\end{equation}
while
\begin{equation}
h=\textrm{e}^{ig\,\xi^{a}T^{a}},
\label{hxi}
\end{equation}
Now, the local gauge invariance  of $A^{h}_{\mu}$ under a gauge transformation $u\in SU(N)$ can be obtained from
\begin{equation}
h\to u^\dagger h\,,\qquad\ h^\dagger\to h^\dagger u\,,\qquad A_\mu \to u^\dagger A_\mu u + \frac{i}{g}u^\dagger \p_\mu u.
\end{equation}
Now, considering the BRST invariance, the Gribov-Zwanziger action in the linear covariant gauges, the total action is given by
\begin{equation}\label{action}
S= S_{YM} + S_{GF} + S_{GZ} + S_{\varepsilon} ,
\end{equation}
whereby $S_{YM}$ is the Yang-Mills action,
 \begin{equation}
S_{YM}  = \frac{1}{4} \int d^4x  F^a_{\mu\nu} F^a_{\mu\nu}   , \label{ym}
\end{equation}
$S_{GF}$ is the Faddeev-Popov gauge-fixing in linear covariant gauges,
\begin{equation}
S_{GF} = \int d^{4}x  \left( \frac{\alpha}{2}\,b^{a}b^{a}
+ib^{a}\,\partial_{\mu}A^{a}_{\mu}
+\bar{c}^{a}\partial_{\mu}D^{ab}_{\mu}(A)c^{b}     \right)   , \label{sfpb}
\end{equation}
$\alpha$ being the gauge parameter and $\alpha=0$ in Landau gauge; $S_{GZ}$ is the Gribov-Zwanziger action in its local form,
\begin{eqnarray}
S_{GZ}&=&\int d^{4}x\left[\bar{\varphi}^{ac}_{\mu}\partial_{\nu}D^{ab}_{\nu}(A^h)\varphi^{bc}_{\mu}-\bar{\omega}^{ac}_{\mu}\partial_{\nu}(D^{ab}_{\nu}(A^h)\omega^{bc}_{\mu})\right]
\nonumber\\&&-\gamma^{2}g\int d^4 x\left[f^{abc}(A^h)^{a}_{\mu}\varphi^{bc}_{\mu}+f^{abc}(A^h)_{\mu}^{a}\bar{\varphi}_{\mu}^{bc}+\frac{d}{g}(N_{c}^{2}-1)\gamma^{2}\vphantom{\frac{1}{2}}\right],
\label{S_gamma2}
\end{eqnarray}
with $(\bar{\varphi}^{ac}_{\mu},~\varphi^{ac}_{\mu})$ a pair of complex-conjugate bosonic fields, $(\bar{\omega}^{ac}_{\mu},~\omega^{ac}_{\mu})$ a pair of anti-commuting complex-conjugate fields; and $\gamma$ the Gribov parameter which is dynamically fixed by a gap equation that gives us the horizon function \cite{Zwanziger:1989mf,Zwanziger:1992qr},
 \begin{equation}\label{gap}
\braket{f^{abc}(A^h)_\mu^a({\varphi}_{\mu}^{bc}+\bar{\varphi}_{\mu}^{bc})}=2d(N^2-1)\frac{\gamma^2}{g^2},
\end{equation}
which can also be rewritten as \cite{Dudal:2011gd}
\begin{equation}\label{gapG}
\frac{\p \Gamma}{\p \gamma^2}=0,
\end{equation}
whereby $\Gamma$ is the quantum action defined by
\begin{equation}\label{quantumAction}
\textrm{e}^{-\Gamma}=\int[d\Phi ] \textrm{e}^{-S}.
\end{equation}
The last term from \eqref{action},
\begin{equation}
S_{\varepsilon} = \int d^4x\; \varepsilon^{a}\,\partial_{\mu}(A^h)^{a}_{\mu}   \label{stau}
\end{equation}
ensures, through the Lagrange multiplier $\varepsilon$, the transversality of the composite operator $(A^h)_\mu^a$, $\partial_{\mu}(A^h)^{a}_{\mu}=0$.

The action $S$, \eqref{action}, enjoys an exact BRST invariance, $s S = 0$ and $s^2=0$ \cite{Capri:2015ixa}
\begin{eqnarray}
&&s A^{a}_{\mu}=-D^{ab}_{\mu}c^{b}\,,\;\;\;\; \;\;\;\; s c^{a}=\frac{g}{2}f^{abc}c^{b}c^{c}\,,\nonumber\\
&&s \bar{c}^{a}=ib^{a}\,,\;\;\;\;\;\;\;\;\;\;\;\;\;\;\;\;\;
s b^{a}= 0\,,  \nonumber\\
&&s \varphi^{ab}_{\mu}= 0 \,,\;\;\;\;\;\;\;\;\;\;\;\;\;\;\;\;\;\;  s \omega^{ab}_{\mu}=0\,,\nonumber\\
&&s\bar\omega^{ab}_{\mu}=0  \,,\;\;\;\;\;\;\;\;\;\;\;\;\;\;\;\;\;\;  s\bar\varphi^{ab}_{\mu}=0\,,\nonumber\\
&&s\varepsilon^{a}=0    \,,\;\;\;\;\;\;\;\;\;\;\;\;\;\;\;\;\;\;\;\;\;   s (A^h)^{a}_{\mu}= 0 \;, \nonumber\\
&&s h^{ij} = -ig c^a (T^a)^{ik} h^{kj}.
 \label{brstgamma}
\end{eqnarray}

\section{Refined Gribov-Zwanziger Action}\label{GZ action}

The BRST invariant $d=2$ condensates, $\langle A_{h,\mu}^{a}A_{h,\mu}^{a}\rangle$ and $\langle \bar{\varphi}^{ab}_{\mu}\varphi^{ab}_{\mu} \rangle$, cause non-perturbative dynamical instabilities disturbing the Gribov-Zwanziger formalism \cite{Dudal:2007cw,Dudal:2008sp,Dudal:2011gd}.
 $\langle \bar{\varphi}^{ab}_{\mu}\varphi^{ab}_{\mu} \rangle$ guarantees that the gluon propagator is non-vanishing at zero momentum, and $\langle A_{\mu}^{a}A_{\mu}^{a}\rangle$ is crucial to fit the lattice data \cite{Dudal:2011gd}. The refined Gribov-Zwanziger action (RGZ) is obtained adding these condensates to the GZ action via the {\it local composite operator (LCO) formalism}, see \cite{Dudal:2011gd}. The operators $A^hA^h$ and $\bar{\varphi}\varphi$ will be added to the action via two BRST invariant bosonic sources $\tau$ and $Q$,
\begin{equation}
s\tau=0 \;\;\;\;\;\mathrm{and}\;\;\;\; sQ=0.
\end{equation}
From here, we opted for the Landau gauge $\p A=0$ for convenience, so that we can work with $A^h=A$, as formally proven in \cite{Capri:2015ixa}.
Then, the action with these operators is written as
\begin{equation}
  \Sigma=S+S_{A^2}+S_{\varphi\bar{\varphi}}+S_{\rm{vac}},
  \label{action_condensate}
\end{equation}
whereby $S$ is given by \eqref{action}
and we also have
\begin{align}
S_{A^2}&=\int d^d x\frac{\tau}{2}A^a_\mu A^a_\mu, \nonumber\\
S_{\bar{\varphi}\varphi}&=\int d^d x Q\bar{\varphi}^{ac}_\mu \varphi^{ac}_\mu,\nonumber\\
S_{\rm{vac}}&=-\int d^d x \left(\frac{\zeta}{2}\tau^2+\alpha QQ +\chi Q\tau\right).
\label{actions}
\end{align}
The parameters $\alpha,\, \chi$ and $\zeta$ are the LCO parameters which guarantee that the divergences of the kind $\Braket{A^2(x)A^2(y)}_{x\rightarrow y}$, etc. can be properly dealt with, see \cite{Dudal:2011gd}.

\section{The Effective Action Calculus} \label{work}
In order to get the effective action, we have written the energy functional as
\begin{equation}\label{W}
e^{-W(J)} = \int [d\Phi] e^{-\Sigma},
\end{equation}
where $\Sigma$ is given by \eqref{action_condensate}.

The action $\Sigma$, \eqref{action_condensate}, has
three terms quadratic in the sources and they should be removed to facilitate calculations and interpret our results in terms of the vacuum energy.
The easiest way to remove these terms is by introducing two auxiliary fields $\sigma_1$ and $\sigma_2$ via two identities written in terms of renormalized fields 
\begin{eqnarray}\label{HS transfo}
1 &=& \int [\mathcal{D} \sigma_1] \ e^{- \frac{1}{2} \int d^d x \left(\sigma_1 + \frac{\bar{a}}{2} A^2 + \bar{b} Q + \bar{c}\tau \right)^2},\nonumber\\
1 &=& \int [\mathcal{D} \sigma_2] \ e^{+ \frac{1}{2} \int d^d x \left(\sigma_2 + \bar{d} \overline{\varphi} \varphi + \bar{e} Q + \frac{\bar{f}}{2} A^2 \right)^2},
\end{eqnarray}
multiplying the integral in \eqref{W}.
If we choose in dimensional regularization ($d=4-\epsilon/2$)
\begin{eqnarray}
\bar{a} &=& - \frac{Z_A}{\sqrt{Z_{\zeta \zeta} \zeta}} \mu^{\epsilon/2}
\nonumber\\
\bar{b} &=&  \frac{Z_{QQ}Z_{\chi \chi} \chi}{\sqrt{Z_{\zeta \zeta} \zeta}} \mu^{- \epsilon/2},
\nonumber\\
\bar{c} &=&  Z_{\tau \tau} \sqrt{Z_{\zeta \zeta} \zeta} \mu^{- \epsilon/2},
\nonumber\\
\bar{d} &=&  \frac{Z_{\varphi}}{\sqrt{\frac{Z_{\chi \chi}^2 \chi^2}{Z_{\zeta \zeta} \zeta}- 2 Z_{\alpha \alpha}\alpha}}\mu^{\epsilon/2},\\
\bar{e} &=&  Z_{QQ} \sqrt{\frac{Z_{\chi \chi}^2 \chi^2}{Z_{\zeta \zeta} \zeta}- 2 Z_{\alpha \alpha}\alpha} \ \mu^{-\epsilon/2},
\nonumber\\
\bar{f} &=& \frac{Z_A}{\sqrt{Z_{\zeta\zeta}\zeta}}\left(\frac{Z_{\tau Q}Z_{\zeta\zeta}\zeta-Z_{QQ}Z_{\chi\chi}\chi}{Z_{QQ}\sqrt{Z_{\chi\chi}^2\chi^2-2Z_{\alpha\alpha}Z_{\zeta\zeta}\zeta\alpha}}\right)\mu^{\epsilon/2},\nonumber
\end{eqnarray}
we can remove the quadratic terms in sources.
In the $\overline{\rm{MS}}$ scheme and at one loop, the $Z$ factors are given by \cite{Dudal:2011gd}
\begin{eqnarray}\label{Zs}
Z_A&=&1+\frac{13}{3}\frac{Ng^2}{16\pi^2\epsilon}, \;\;\;\;\;\;\;\;\;\;\;
\tilde Z_{\zeta} = Z_{\zeta\zeta} Z_{\tau\tau}^2 = 1-\frac{13}{3}\frac{Ng^2}{16\pi^2\epsilon}, \;\;\;\;\;\;\;\;\;\;\;\;\;\;\;
Z_{\zeta\zeta} = 1+\frac{22}{3}\frac{Ng^2}{16\pi^2\epsilon},\nonumber\\
Z_g&=&1-\frac{11}{3}\frac{Ng^2}{16\pi^2\epsilon},\;\;\;\;\;\;\;\;\;\;\;
Z_{\tau}=1-\frac{35}{6}\frac{Ng^2}{16\pi^2\epsilon},\;\;\;\;\;\;\;\;\;\;\;\;\;\;\;
Z_{QQ}=Z_gZ_{A}^{1/2}=1-\frac{3}{2}\frac{Ng^2}{16\pi^2\epsilon},\nonumber\\
Z_{\chi\chi}&=&1, \;\;\;\;\;\;\;\;\;\;\;\;\;\;\;\;\;\;\;\;\;\;\;\;\;\;\;\;\;\;\;\;\;
Z_{\tau Q}=0,\;\;\;\;\;\;\;\;\;\;\;\;\;\;\;\;\;\;\;\;\;\;\;\;\;\;\;\;
\tilde Z_{\alpha}=Z_{\alpha\alpha}Z_{QQ}^2=1+\frac{35}{6}\frac{Ng^2}{16\pi^2\epsilon},\nonumber\\
Z_{\alpha\alpha}&=&1+\frac{53}{6}\frac{Ng^2}{16\pi^2\epsilon},\;\;\;\;\;\;\;\;\;\;\;\;\;\;\;\;\;\;\;\;\;\;
Z_{\varphi}=Z_{\bar{\varphi}}=Z_g^{-1}Z_A^{-1/2}=1+\frac{3}{2}\frac{Ng^2}{16\pi^2\epsilon}.
\end{eqnarray}
Therefore, \eqref{W} becomes
\begin{eqnarray}\label{W1}
e^{-W(Q,\tau)}=
\int [\mathcal{D}\Phi] [\mathcal{D}\sigma_{1,3}] \exp &&\left[-S_{GZ}-\frac{1}{2} \int d^dx \left(2\bar{c} \sigma_1 \tau + 2 \sigma_3  Q
\left(1-\frac{\bar{b}^2}{\bar{e}^2} \right)\sigma_1^2\right.\right.\nonumber\\
&&\left.\left. - \frac{1}{\bar{e}^2} (\sigma_3^2-2\bar{b}\sigma_1\sigma_3)+ \left(\left(\bar{a} -\frac{\bar{f}\bar{b}}{\bar{e}}\right)\Braket{\sigma_1}+\frac{\bar{f}}{\bar{e}}\Braket{\sigma_3}\right) A^2 \right.\right.\nonumber\\
&&\left.\left.
- 2 \frac{\bar{d}}{\bar{e}}(\bar{b}\Braket{\sigma_1}-\Braket{\sigma_3}) \overline{\varphi} \varphi \right)\right],
\end{eqnarray}
where
\begin{equation}\label{sigma3}
\sigma_3=\sigma_1 \bar{b} - \sigma_2 \bar{e}.
\end{equation}
So far, all LCO parameters, sources and fields have been renormalized, except the auxiliary fields $\sigma$'s.  Analyzing the term $\bar{c} \sigma_1 \tau = Z_{\tau} \sqrt{Z_{\zeta} \zeta} \mu^{- \epsilon/2} \sigma_1 \tau$ that appears in Eq.\eqref{W1}, it is easy to see that this field is indeed infinite, thus it also must be renormalized. As the $Z$-factors are infinite, if $\sigma_1$ would be finite, then the quantity multiplying $\tau$ would be infinite. This would not make sense as a physical (and thus finite) local operator $\mathcal O$. The original $\sigma_1$ field \emph{should} then be infinite in order to get a finite quantity multiplying the finite source $\tau$. It is then natural to define a renormalized finite field $\sigma_1'$ by $\sigma_1' \equiv Z_{\tau \tau} \sqrt{Z_{\zeta \zeta}} \sigma_1 \equiv \sqrt{{\widetilde Z}_\zeta} \sigma_1$. Regarding $\sigma_3$, the term $\sigma_3Q$ in Eq.\eqref{W1} teaches us, knowing that $Q$ is finite, that $\sigma_3$ is already finite and thus should not be renormalized. In terms of the \emph{finite} fields $\sigma_1'$ and $\sigma_3$, the energy functional now reads :
\begin{eqnarray}\label{W2}
e^{-W(Q,\tau)} S= \int [\mathcal{D}\Phi] [\mathcal{D}\sigma_{1,3}] \ \exp && \left[-S_{GZ}-\frac{1}{2} \int d^dx \left(\vphantom{\frac{\sigma_1'}{\sqrt{{\widetilde Z}_\zeta}}}-2\sqrt{\zeta} \sigma_1' \tau + 2 \sigma_3  Q \left(1-\frac{\bar{b}^2}{\bar{e}^2} \right)\frac{{\sigma_1'}^2}{{\widetilde Z}_\zeta} \right.\right.\nonumber\\
&&\left.\left. - \frac{1}{\bar{e}^2} \left(\sigma_3^2-2\bar{b}\frac{\sigma_1'}{\sqrt{{\widetilde Z}_\zeta}}\sigma_3\left(\bar{a} -\frac{\bar{f}\bar{b}}{\bar{e}}\right)\frac{\Braket{\sigma_1'}}{\sqrt{{\widetilde Z}_\zeta}}+\frac{\bar{f}}{\bar{e}}\Braket{\sigma_3}\right) A^2 \right.\right.\nonumber\\
&&\left.\left.
- 2 \frac{\bar{d}}{\bar{e}}\left(\bar{b}\frac{\Braket{\sigma_1'}}{\sqrt{{\widetilde Z}_\zeta}}-\Braket{\sigma_3}\right) \overline{\varphi} \varphi \right)\vphantom{\frac{\sigma_1'}{\sqrt{{\widetilde Z}_\zeta}}}\right].
\end{eqnarray}
In this expression, all LCO parameters, sources and fields are now finite, and infinities are only present in the renormalization factors $Z$'s, explicitly written or present in the coefficients $\bar a,...,\bar f$.
At one loop, $\chi=0$, $Z_{\tau Q} = 0$ \cite{Dudal:2011gd} which implies that $\bar{b}=\bar{f}=0$, then $\sigma_3=-\bar{e}\sigma_2$. Now, by analysis of the term $\sigma_3Q=-\bar{e}\sigma_2Q$ in \eqref{W2}, a similar reasoning as above for $\sigma_1$ shows that $\sigma_2$ is infinite and should be renormalized defining a new finite field $\sigma_2'$ through $\sigma_2' \equiv Z_{QQ} \sqrt{Z_{\alpha \alpha}} \sigma_2 \equiv \sqrt{\widetilde{Z_{\alpha}}}\sigma_2$. Hence, the energy functional in terms of the finite fields $\sigma_1'$ and $\sigma_2'$ and with one-loop coefficients reads :
\begin{eqnarray}\label{Final W}
e^{-W(Q,\tau)} = \int \vphantom{\frac{{\sigma_1'}^2}{{\widetilde Z}_\zeta}}[\mathcal{D}\Phi] [\mathcal{D}\sigma_{1,2}]  \exp && \left[\vphantom{\frac{{\sigma_1'}^2}{{\widetilde Z}_\zeta}}-S_{GZ}
-\frac{1}{2} \int \vphantom{\frac{{\sigma_1'}^2}{{\widetilde Z}_\zeta}} d^dx \left(\frac{{\sigma_1'}^2}{{\widetilde Z}_\zeta} - \frac{{\sigma_2'}^2}{{\widetilde Z}_\alpha}
+ \bar{a}\frac{\Braket{\sigma_1'}}{\sqrt{{\widetilde Z}_\zeta}} A^2
- 2 \bar{d}\frac{\Braket{\sigma_2'}}{\sqrt{{\widetilde Z}_\alpha}} \overline{\varphi} \varphi\right.\right.\nonumber\\
&&\left.\left. -2\sqrt{\zeta} \sigma_1' \tau + 2 \sqrt{-2\alpha} \sigma_2'  Q \vphantom{\frac{{\sigma_1'}^2}{{\widetilde Z}_\zeta}}\right)\vphantom{\frac{{\sigma_1'}^2}{{\widetilde Z}_\zeta}}\right]
\end{eqnarray}
In this expression, infinities are now localized, only in the renormalization factors $\tilde Z_{\zeta}$, $\tilde Z_{\alpha}$, and in those hidden in $\bar{a}$ and $\bar{d}$.

In order to have an expression of the form $\frac{m^2}{2}A^2-M^2\bar{\varphi}\varphi$, we defined the effective masses, $m^2$ and $M^2$, respectively linked to $\Braket{AA}$ and $\Braket{\bar{\varphi}\varphi}$ by:
\begin{equation}\label{m^2 infinite}
m^2 \equiv \frac{\bar{a}}{\sqrt{\widetilde{Z_{\zeta}}}} \Braket{\sigma_1'}
=\left(1+\frac{17}{6} \frac{Ng^2}{16\pi^2\epsilon}\right) \sqrt{\frac{13Ng^2}{9(N^2-1)}}\Braket{\sigma_1'}+\mathcal{O}(g^4),
\end{equation}
\begin{equation}\label{M^2 infinite}
M^2 \equiv \frac{\bar{d}}{\sqrt{\widetilde{Z_{\alpha}}}} \Braket{\sigma_2'}
=-\left(1-\frac{35}{6} \frac{Ng^2}{16\pi^2\epsilon} \right)\sqrt{\frac{35Ng^2}{48(N^2-1)^2}} \Braket{\sigma_2'}+\mathcal{O}(g^4).
\end{equation}
where the last equalities follow from considering the first order term of the $Z$-factors in $\overline{\rm{MS}}$ scheme and $\alpha=\frac{\alpha_0}{g^2}=-\frac{24(N^2-1)^2}{35Ng^2}$ and $\zeta=\frac{\zeta_0}{g^2}=\frac{9(N^2-1)}{13Ng^2}$ \cite{Dudal:2011gd}. 

The ghost fields $c$, $\bar{c}$, $\omega$, $\bar \omega$ give us just an overall factor. Now, to integrate over the $\varphi$ and $\bar{\varphi}$ fields, we proceed as follows
\begin{equation}
\bar{\varphi}_{\mu}^{ab}=U_{\mu}^{ab}+iV_{\mu}^{ab},~~\varphi_{\mu}^{ab}=U_{\mu}^{ab}-iV_{\mu}^{ab}.
\end{equation}
Then
\begin{eqnarray}
\int[\mathcal{D}U,V]&&e^{-\int d^dx \left[V_{\mu}^{ab}P_{\mu\nu}^{ac,bd}V_{\nu}^{cd}+U_{\mu}^{ab}P_{\mu\nu}^{ac,bd}U_{\nu}^{cd}-2g\gamma^2f^{abc}A_{\mu}^{a}U_{\mu}^{bc}\right]}=\nonumber\\
&=&\frac{1}{\det P_{\mu\nu}^{ac,bd}}e^{\int d^dx Ng^2\gamma^4A_{\mu}^aP_{\mu\nu}^{-1}\delta^{ab}A_{\nu}^{b}},
\end{eqnarray}
with
\begin{equation}
P_{\mu\nu}^{ac,bd}\equiv(\partial^2-M^2)\delta^{ac}\delta^{bd}\delta_{\mu\nu}.
\end{equation}
Therefore, the first contribution $\Gamma_a$ to the effective potential is obtained by :
\begin{equation}
\Omega\Gamma_a = \ln\det P_{\mu\nu}^{ac,bd} = {\rm Tr}\ln P_{\mu\nu}^{ac,bd},
\end{equation}
resulting in
\begin{eqnarray}\label{gammaa}
\Gamma_a &=& (N^2-1)^2 \left[- \frac{1}{\epsilon}\frac{M^4}{4\pi^2} + \frac{M^4}{8\pi^2} \ln \frac{M^2}{\overline{\mu}^2}-\frac{M^4}{8\pi^2} \right] \nonumber\\
&=& \frac{35Ng^2}{48}\frac{\Braket{\sigma_2'}^2}{4\pi^2} \left(-\frac{1}{\epsilon}-\frac{1}{2}+\frac{1}{2} \ln\left(- \sqrt{\frac{35Ng^2}{48(N^2-1)^2}}\frac{\Braket{\sigma_2'}}{\bar{\mu}^2} \right)\right)+\mathcal{O}(g^4).
\end{eqnarray}
The second contribution $\Gamma_b$ to the effective potential comes from the gluon field $A_{\mu}$. The quadratic part of the action containing $A_{\mu}$ is
\begin{equation}
e^{-\frac{1}{2}\int d^dx A_{\mu}^aR_{\mu\nu}^{ab}A_{\nu}^b}
\end{equation}
where
\begin{equation}
R_{\mu\nu}^{ab}\equiv\delta^{ab}\left[\left(-\partial^2+m^2-\frac{2N\gamma^4g^2}{\partial^2-M^2} \right)\delta_{\mu\nu}-\partial_{\mu}\partial_{\nu}\left(\frac{1}{\alpha}-1\right)\right].
\end{equation}
Therefore,
\begin{equation}
\Omega\Gamma_b = \frac{1}{2}\ln\det R_{\mu\nu}^{ac,bd} = \frac{1}{2}{\rm Tr}\ln R_{\mu\nu}^{ac,bd},
\end{equation}
resulting in
\begin{eqnarray}\label{Gammab}
\Gamma_b&=&-\frac{(N^2-1)}{2(4\pi)^2}\left(\frac{3}{\epsilon}{+\frac{5}{4}}\right)\left(m^4-4\gamma^4g^2N\right)+
\frac{3(N^2-1)}{4(4\pi)^2}\left[x_1^2\ln\left(\frac{-x_1}{\bar{\mu}^2}\right)+x_2^2\ln\left(\frac{-x_2}{\bar{\mu}^2}\right)\right.\nonumber\\
&&\left.-M^4\ln\left(\frac{M^2}{\bar{\mu}^2}\right)\right],
\end{eqnarray}
where $x_1$ and $x_2$ are the solutions of the equation $x^2+(M^2+m^2)x+M^2m^2+\lambda^4=0$,
\begin{eqnarray}
x_1&=&-\frac{1}{2}\left(m^2+M^2+\sqrt{(m^2-M^2)^2-4\lambda^4}\right),\nonumber\\
x_2&=&-\frac{1}{2}\left(m^2+M^2 - \sqrt{(m^2-M^2)^2+-4\lambda^4}\right).
\end{eqnarray}
The third part of effective potential $\Gamma_c$ is from the Gribov-Zwanziger action,
\begin{equation}
\Gamma_c=-d\gamma_0^4(N^2-1).
\end{equation}
Knowing that $Z_{\gamma^2}=Z_g^{-1/2}Z_A^{-1/4}$, we get
\begin{equation}
\gamma_0^4=Z^2_{\gamma^2}\gamma^4,\;\;\; {\rm with}\;\;\;  Z^2_{\gamma^2}=1+\frac{3}{2}\frac{g^2N}{16\pi^2\epsilon},
\end{equation}
hence
\begin{equation}\label{gammac}
\Gamma_c
= (N^2-1)\gamma^4\left(-4+\frac{3Ng^2}{32\pi^2}-\frac{3Ng^2}{8\pi^2\epsilon}\right)+\mathcal{O}(g^4).
\end{equation}
And the last contribution comes from $\Braket{\sigma_1'}^2$ and $\Braket{\sigma_2'}^2$ :
\begin{eqnarray}\label{gammad}
\Gamma_d &=& \frac{1}{2} \left(\frac{1}{\tilde Z_{\zeta}} \Braket{\sigma_1'}^2 - \frac{1}{\tilde Z_{\alpha}} \Braket{\sigma_2'}^2 \right)\nonumber\\
&=& \frac{\Braket{\sigma_1'}^2}{2} - \frac{\Braket{\sigma_2'}^2}{2}+\frac{13}{6} \frac{Ng^2}{16\pi^2\epsilon}\Braket{\sigma_1'}^2+\frac{35}{12} \frac{Ng^2}{16\pi^2\epsilon}\Braket{\sigma_2'}^2+\mathcal{O}(g^4).
\end{eqnarray}
The full effective potential given by $\Gamma=\Gamma_a+\Gamma_b+\Gamma_c+\Gamma_d$ is finite when $\epsilon \rightarrow 0$ at first order in $g^2$. Therefore, it can be written as
\begin{multline}\label{effective potential}
\Gamma(m^2,M^2,\lambda^4)=\frac{9(N^2-1)}{13Ng^2}\frac{m^4}{2}-\frac{48(N^2-1)^2}{35Ng^2}\frac{M^4}{2}-\frac{2\lambda^4(N^2-1)}{Ng^2}\\
-\frac{N^2-1}{16\pi^2}\left\{-2\lambda^4+\frac{5}{8}m^4+2(N^2-1)M^4-\left(2(N^2-1)-\frac{3}{4}\right)\ln\left(\frac{M^2}{\bar{\mu}^2}\right)M^4\right\}\\
+\frac{3}{8}\frac{N^2-1}{16\pi^2}\left\{m^4+M^4-2\lambda^4
+\left(m^2+M^2\right)\sqrt{(m^2-M^2)^2-4\lambda^4}\right\} \\
\times \ln\left[\frac{1}{2\bar{\mu}^2}\left(m^2+M^2+\sqrt{(m^2-M^2)^2-4\lambda^4}\right)\right]\\
+\frac{3}{8}\frac{N^2-1}{16\pi^2}\left\{m^4+M^4-2\lambda^4
-\left(m^2+M^2\right)\sqrt{(m^2-M^2)^2-4\lambda^4}\right\} \\
\times \ln\left[\frac{1}{2\bar{\mu}^2}\left(m^2+M^2-\sqrt{(m^2-M^2)^2-4\lambda^4}\right)\right].
\end{multline}
with
\begin{eqnarray}
m^2 &=& \sqrt{\frac{13Ng^2}{9(N^2-1)}} \Braket{\sigma_1'},\\
M^2 &=& - \sqrt{\frac{35Ng^2}{48(N^2-1)^2}} \Braket{\sigma_2'}.
\end{eqnarray}
and $\lambda^4 \equiv 2Ng^2\gamma^4$.
 The next step is to analyze the gap equations given by
\begin{equation}\label{gapEq}
\frac{\p \Gamma}{\p M^2}=0, \;\;\;\;\;\;\;\; \frac{\p \Gamma}{\p m^2}=0,  \;\;\;\;\;\;\;\;\ \frac{\p \Gamma}{\p \lambda^4}=0.
\end{equation}
Unfortunately, an acceptable result was not obtained in this particular scheme. The resolution was to not fix a scheme, we rather rewrote the effective potential in a general scheme as has been done in e.g.~\cite{Dudal:2005na}. Details of this procedure will be published elsewhere. The effective potential \eqref{effective potential}, in general scheme, becomes
\begin{multline}\label{general effective potential}
\Gamma_{gen}(m^2,M^2,\lambda^4,b_0)=\frac{9(N^2-1)}{26Ng^2}m^4-\frac{24(N^2-1)^2}{35Ng^2}M^4-\frac{2(N^2-1)^2M^4}{16\pi^2}\left(1-\ln\left(\frac{M^2}{\bar{\mu}^2}\right)\right)\\
-2\lambda^4\frac{N^2-1}{Ng^2}
-2\lambda^4\frac{N^2-1}{16\pi^2}( b_0-1 )\\
+\frac{3}{4}\frac{N^2-1}{16\pi^2}\left\{\frac{5}{4}(m^4+M^4-2\lambda^4)-
\frac{m^2+M^-2\lambda^4}{2}\ln\left[\frac{m^2M^2+\lambda^4}{\bar{\mu}^4}\right]\right.\\
\left.+(m^2+M^2)\sqrt{4\lambda^4-(m^2-M^2)^2}\arctan\left[\frac{\sqrt{4\lambda^4-(m^2-M^2)^2}}{m^2+M^2}\right]+\ln\left[\frac{M^2}{\bar{\mu}^2}\right]M^4
\right\},
\end{multline}
$b_0$ being a parameter related to the chosen scheme for the coupling. It was fixed, at the end, by matching our values for the complex conjugate poles masses of the transverse gluon propagator to those estimated from lattice data \cite{Dudal:2018cli} when the gap equations are solved for. The effective masses $m^2$ and $M^2$ and the Gribov parameter $\gamma^2$ were obtained as functions of the parameter $b_0$ and $\bar{\mu}$ in units $\Lambda=1$, $N=3$ and also for $N=2$. Notice that these poles masses are gauge and scheme independent \cite{Capri:2015ixa}, so we benefitted from this to fix the parameters $b_0$ and $\bar{\mu}$ by using a minimal external lattice input to determine the ``optimum scheme''. We got $b_0=-3.643$ and $\bar{\mu}=1.429$. With this procedure, we obtained a reasonable value for the coupling constant, namely $0.382$. Therefore, in this case, the perturbative result is relatively trustworthy. The Gribov parameter $\gamma^2$ is $0.637$ and the vacuum energy is $-26.955$. The Hessian determinant is positive and also the second derivatives,
\begin{equation}\label{2derivative}
\left.\frac{\p^2 \Gamma_{gen}}{\p {M^2}^2}\right|_{solved}=1.668 \;\;\;\;\;\;\;\; \left.\frac{\p^2 \Gamma_{gen}}{\p {m^2}^2}\right|_{solved}=0.216  \;\;\;\;\;\;\;\;\ \left.\frac{\p^2 \Gamma_{gen}}{\p M^2\p m^2}\right|_{solved}= 0.011.
\end{equation}
Then, the solution does correspond to a minimum.

The future step will be to extend this research to finite temperatures and  to study if the deconfinement transitions reflects itself in a change in the propagator behavior and to check if, with the Polyakov loop added to the game, we observe the transition also in that order parameter.


%
%
%

\end{document}